\documentclass{article}

\usepackage{arxiv}

\usepackage[utf8]{inputenc} 
\usepackage[T1]{fontenc}    
\usepackage{hyperref}       
\usepackage{url}            
\usepackage{booktabs}       
\usepackage{amsfonts}       
\usepackage{nicefrac}       
\usepackage{microtype}      
\usepackage{lipsum}
\usepackage{graphicx}
\usepackage{amssymb}
\usepackage{lipsum}
\usepackage{tabularray}
\usepackage[linesnumbered,ruled]{algorithm2e}
\usepackage{booktabs}
\usepackage{multirow}
\usepackage{siunitx}
\usepackage{mhchem}
\usepackage[numbers]{natbib}
\graphicspath{ {./images/} }

\title{Neutron particle transport  3D method of characteristic   Multi GPU platform  Parallel Computing}

\author{
Faguo Zhou\\
School of Artificial Intelligence\\ 
China University of Mining and Technology-Beijing \\
Beijing 100083, China\\
\texttt{ } 
   \And
 Shunde Li \\
  Computer Network Information Center\\
  Chinese Academy of Sciences\\ 
  Beijing 100083, China\\
  \texttt{ } 
  \And
 Rong Xue \\
School of Artificial Intelligence\\ 
China University of Mining and Technology-Beijing \\
Beijing 100083, China\\
  \texttt{ } \\  
    \And
 Lingkun Bu \\
  National Center for Materials Service Safety\\
  University of Science and Technology Beijing\\
  Beijing 100083, China\\
  \texttt{ } \\
    \And
 Ningming Nie \\
  Computer Network Information Center\\
  Chinese Academy of Sciences\\ 
  Beijing 100083, China\\
  \texttt{ } 
    \And
 Peng Shi \thanks{Corresponding author } \\
  National Center for Materials Service Safety\\
  University of Science and Technology Beijing\\
  Beijing 100083, China\\
\texttt{pshi@ustb.edu.cn}\\
    \And
 Jue Wang \thanks{Corresponding author } \\
 Computer Network Information Center\\
  Chinese Academy of Sciences\\ 
  Beijing 100083, China\\
   \texttt{wangjue@sccas.cn} \\
    \And
 Yun Hu \\
  China Institute of Atomic Energy\\
  Beijing 102413, China
  \texttt{ } \\
    \And
 Zongguo Wang \\
 Computer Network Information Center\\
  Chinese Academy of Sciences\\ 
  Beijing 100083, China\\
  \texttt{ } \\
    \And
 Yangang Wang \\
 Computer Network Information Center\\
  Chinese Academy of Sciences\\ 
  Beijing 100083, China\\
  \texttt{ } \\
  \And
 Qinmeng Yang \\
 Computer Network Information Center\\
  Chinese Academy of Sciences\\ 
  Beijing 100083, China\\
  \texttt{ } \\
    \And
 Miao Yu \\
School of Artificial Intelligence\\ 
China University of Mining and Technology-Beijing \\
Beijing 100083, China\\
  \texttt{ } \\  
  \texttt{ } \\
}

\begin{document}
\maketitle
\begin{abstract}
Three-dimensional neutron transport calculations using the Method of Characteristics (MOC) are highly regarded for their exceptional computational efficiency, precision, and stability. Nevertheless, when dealing with extensive-scale computations, the computational demands are substantial, leading to prolonged computation times. To address this challenge while considering GPU memory limitations, this study transplants the real-time generation and characteristic line computation techniques onto the GPU platform. Empirical evidence emphasizes that the GPU-optimized approach maintains a heightened level of precision in computation results and produces a significant acceleration effect. Furthermore, to fully harness the computational capabilities of GPUs, a dual approach involving characteristic line preloading and load balancing mechanisms is adopted, further enhancing computational efficiency. The resulting increase in computational efficiency, compared to traditional methods, reaches an impressive 300 to 400-fold improvement.
\end{abstract}


\section{Introduction}

In recent years, Numerical Reactor becomes a key technology for simulating and analyzing nuclear reactors using numerical methods \citep{mccaskey2011nuclear}. By using numerical methods on computers to solve the physics and thermal-hydraulic problems of nuclear reactors \citep{chauliac2011nuresim}, Numerical Reactor aims to simulate the physical processes and coupled effects inside the reactor to obtain detailed information about reactor performance, thermal behavior, material behavior, and radiation characteristics, providing support for reactor design, optimization, and safety analysis. In this regard, developed countries in Europe and America initiate research projects on Numerical Reactor in the past two decades \citep{deng2016key} and conduct numerous simulation experiments on operating nuclear reactors using supercomputers.In virtual reactors, the biggest problem is obtaining a exact solution of neutron transport equation.

Neutron transport calculations are one of the crucial functionalities for computing neutron transport behavior in a reactor, providing the required detailed neutron flux density distribution for other simulation processes in the numerical reactor. Among them, the Method of Characteristics (MOC) \citep{askew1972characteristics} has gradually become a research hotspot due to its geometric adaptability, high computational accuracy, a computationally concise and implementable model, and powerful parallel potential. With

The core idea of the MOC method is to parameterize the reactor cross-sections and transform the neutron transport problem into a problem of solving these parameters. By decomposing the reactor cross-sections into adjacent planes, treating the reaction cross-sections on each plane as constants while allowing for variations among the cross-sections on different planes, the MOC method transforms the neutron transport problem on each plane into a problem of solving these parameters, obtaining the neutron transport behavior on the plane. Finally, by connecting the transport results from different planes, the solution for the neutron transport behavior in the entire space is obtained. 

The MOC method offers several advantages, including high computational efficiency \citep{boyd2013massively} and wide applicability, particularly for complex three-dimensional reactor geometries and challenging reactor physics problems. Furthermore, the MOC method demonstrates good scalability and modularity, facilitating its integration with other numerical methods to enhance computational accuracy. However three-dimensional MOC necessitates substantial computational resources, including storage requirements for numerous geometric grids, characteristic lines, and flux data, as well as the associated iterative and floating-point operation counts, leading to relatively lengthy computation times. In recent years, advancements in computer hardware and high-performance computing technologies have made large-scale parallel MOC feasible \citep{liao2014milkyway,fu2016sunway,zhang2022development,zhang2011acceleration}. Gradually, there has been a rise in parallel three-dimensional MOC programs designed for massive parallelization, such as OpenMOC \citep{gunow2019full}, NECP-X \citep{chen2018new}, APOLLO3 \citep{santandrea2017neutron}, and others, which have achieved significant progress in their respective application domains. However, it's important to note that GPU-based three-dimensional MOC programs are still in the development stage, and the full potential of GPU computing for 3D characteristic line solving has yet to be fully explored \citep{wu2003new}.

To further explore the efficiency of the three-dimensional MOC and leverage the interdisciplinary and complementary nature of reactor simulation and high-performance computing, this study is developing a parallel computing program for three-dimensional neutron transport using GPUs based on the three-dimensional MOC. The primary goal is to achieve higher computational efficiency. This paper is organized as follows. First, we introduce the method of characteristics. The second section describes the OTF (On-the-Fly) method for the three-dimensional MOC. Sections 3 and 4 introduce a GPU-based parallel OTF method for the three-dimensional MOC approach. Section 5 provides numerical experimental results on different scales and platforms. Finally, we discuss the prospects of this method and provide suggestions for further research.

\section{Methodology}
\subsection{Review of MOC}
The neutron transport equation (Boltzmann equation) \citep{askew1972characteristics} is a conservation equation used to study the neutron density distribution within a medium. It represents the conservation relationship of neutrons generated through transport, absorption, scattering, and fission reactions during the reaction process. The MOC (Method of Characteristics) \citep{talamo2013numerical} is capable of solving the steady-state transport equation:

\begin{equation}
    \Omega \cdot \bigtriangledown \Psi \left ( r,\Omega ,E \right )+ {\textstyle \sum_{}^{T}} \left (  r,E\right )\Psi \left ( r,\Omega ,E \right )=Q\left ( r,Q,E \right )
\end{equation}

$\Omega$ represents the angular direction vector; $\Psi$  represents the angular neutron flux; r represents the spatial position vector; E represents the neutron energy; Q represents the angular neutron source;  represents the total neutron cross-section.

After discretization in space, angle, and energy groups, the solution of the transport equation is transformed into a set of characteristic line differential equations along different trajectories (hypothetical neutron motion trajectories) after spatial discretization.

\begin{equation}
\frac{\mathrm{d} }{\mathrm{d} s} \phi_{g,n,m} \left ( s \right )  + {\textstyle \sum_{g}^{T}\phi }_{g,n,m}\left (  s\right )=Q_{g,n,m}\left ( s \right )   
\end{equation}

$\phi_{g,n,m} \left ( s \right )$ represents the angular neutron flux in energy group g, with n as the polar angle and m as the azimuthal angle, along the characteristic line trajectory s. ${\textstyle \sum_{g}^{T} }$ represents the total cross-section in energy group g.

Using the flat source approximation, the spatial grid corresponds to flat source regions (FSR). Within the same grid, the neutron source term is constant. The incident neutron angular flux $\phi_{i,g,n,m}^{in}$ can be used as the initial condition to solve for the neutron flux density distribution along a single characteristic line:

\begin{equation}
\phi_{g,n,m,j}^{out}\left ( s \right )= \phi_{g,n,m,j}^{in}\left ( s \right )e^{-s{\textstyle \sum_{g}^{T} }}+\frac{Q_ {g,n,m} }{\sum_{g}^{T}} \left ( 1-e^{-s{\textstyle \sum_{g}^{T} }} \right )
\label{equation3}
\end{equation}

In the equation, j represents the index of the Flat Source Region (FSR), and the total number of FSRs is J.

By integrating equation~\ref{equation3} along the characteristic line, we can obtain the average neutron angular flux along that specific characteristic line.

\begin{equation}
\overline{\phi_{g,n,m,j}}=\frac{Q_{g,n,m}}{\sum_{g}^{T}}+\frac{1}{s_k\sum_{g}^{T}} \sum_{j\in \left \{ 1,\cdot \cdot \cdot ,J \right \} }^{} \left (  \phi_{g,n,m,j}^{in}-\phi_{g,n,m,j}^{out}\right ) 
\end{equation}

$s_k$ represents the truncation length of the characteristic line within the grid. By summing the weighted average neutron angular flux along all directional trajectories within the grid, we can solve for the neutron flux density in that grid. By sequentially solving in the direction of characteristic lines, we obtain the distribution of neutron flux density throughout the entire computational space.

The above describes a typical workflow of the characteristic line transport calculation. To perform characteristic line calculations, it is necessary to obtain the truncation length of the characteristic lines in the spatial grid in advance. Therefore, ray tracing becomes a geometric preprocessing step unique to the MOC. The specific workflow of the characteristic line transport calculation is determined by the chosen ray tracing method.From equations (2) to (4), it can be observed that the characteristic line transport calculation treats a single characteristic line as the smallest computational unit. The characteristic equations on different trajectories are not strongly coupled during the calculation process. This allows for simultaneous ray tracing calculations for multiple characteristic lines. Consequently, the MOC is well-suited for parallel computing.

\subsection{On-the-fly}
In the neutron transport analysis, MOC encompasses a spectrum of ray tracing \citep{filippone1981particle} methodologies, including Modular Ray Tracing(MRT) \citep{shaner2015theoretical}, On-the-fly(OTF) \citep{gunow2016reducing}, Chord Classification Method(CCM) \citep{sciannandrone2016optimized} and modular spatial domain decomposition (SDD) \citep{kochunas2013hybrid}. However, within the context of the distinctive computational attributes inherent to GPU, a judicious decision has been made to adopt the OTF technique. This selection is underpinned by the exceptional computational capabilities exhibited by GPUs in managing extensive numerical computations. The OTF methodology, thoughtfully designed to harness the parallel processing potential of GPUs, entails the real-time generation of ray trajectories. This approach obviates the necessity for substantial memory storage, thereby ameliorating the memory overhead on GPUs and concomitantly amplifying computational efficiency. Noteworthy is the fact that the OTF methodology seamlessly aligns with the efficient logical inference mechanisms intrinsic to GPUs, ensuring the concurrent realization of real-time ray generation and the upholding of computational accuracy. 

In nuclear reactor core analysis, a notable observation is the remarkable resemblance of reactor cores to axially extruded geometries. In this context, an axially extruded geometry refers to a configuration where all radial segments of the reactor geometry are the same. It is important to note that this requirement pertains specifically to the radial geometry and does not impose constraints on the materials used. When working with axially extruded geometry, you can opt to retain exclusively the 2D segments linked to radial crossings. Subsequently, 3D segments can be dynamically generated in real-time using a simple axial mesh. This approach optimizes memory utilization and computational efficiency, allowing for more streamlined and accurate analyses of axially extruded reactor cores.

In the OTF method, 3D segments are dynamically generated based on the information obtained from the initial ray tracing of 2D tracks. This process ensures that every 3D track aligns with a segmented 2D track. By utilizing the initial first positions and polar orientations, it becomes feasible to ascertain the distances from the track to axial intersections and for radial intersections via the 2D segments.

The OTF axial ray tracing is conducted either individually for each 3D track or collectively for the entire z-stack. This adaptable approach offers flexibility in optimizing computational efficiency and memory utilization based on specific requirements and available resources. By dynamically generating 3D segments on-the-fly and efficiently tracing the relevant intersections, the OTF method enhances the accuracy and computational performance of neutron transport simulations in complex geometries, such as those encountered in nuclear reactor cores.

The layout of the 3D tracks offers a systematic approach for ray tracing across the entire z-stack. In this arrangement, every tracks within the z-stack have the same angle $\theta$, projecting onto the identical 2D track, and are equidistant from one another with a constant axial ray spacing denoted as $\bigtriangleup z$.Hence, by defining $z_{0}\left ( 0 \right )$ as the z-coordinate where the bottom track intersects the z-axis at the start of the corresponding 2D track. The axial height $z_{i}$ of the $i^{th}$ bottom track (starting from 0) can be determined as follows:

\begin{equation}
z_{i}\left ( s \right )=z_{0}\left ( 0 \right )+i\bigtriangleup z+s\cot \theta 
\label{equation5}
\end{equation}

Where s represents the length of the 2D track. The inclusion of the axial height $z_{i}$, computed by considering the parameter s along the corresponding 2D track, in conjunction with the 2D track particulars, provides a comprehensive representation of the path and spatial positioning of the 3D tracks within the stack. Consequently, this approach allows us to accurately identify the specific tracks that will pass through a designated Fine Spatial Region (FSR).

For every 2D segment contained within the 2D track, there corresponds an axially extruded region that encompasses a collection of 3D Fine Spatial Regions within that area. While tracing a stack, crossings within these axially extruded Fine Spatial Regions are established in the sequence of traversal by the 2D segments. To establish the axial limits of each 3D FSR, the 1D axial mesh, which can be either local or global and is linked to the axially extruded region, is utilized. By using the FSR boundaries and applying Equation~\ref{equation5}, it becomes feasible to analytically determine the track indices that will intersect the FSR.

\begin{equation}
i_{start}=\left \lceil\frac{z_{min}-max\left ( z_{0}\left ( s_{start}  \right ),z_{0}\left ( s_{end}  \right )     \right )   }{\bigtriangleup z} \right \rceil
\label{equation6}
\end{equation}

\begin{equation}
i_{end}=\left \lfloor \frac{z_{max}-min\left ( z_{0}\left ( s_{start}  \right ),z_{0}\left ( s_{end}  \right )     \right )   }{\bigtriangleup z}  \right \rfloor
\label{equation7}
\end{equation}

\begin{figure}
\centering
\includegraphics[width=0.6\textwidth]{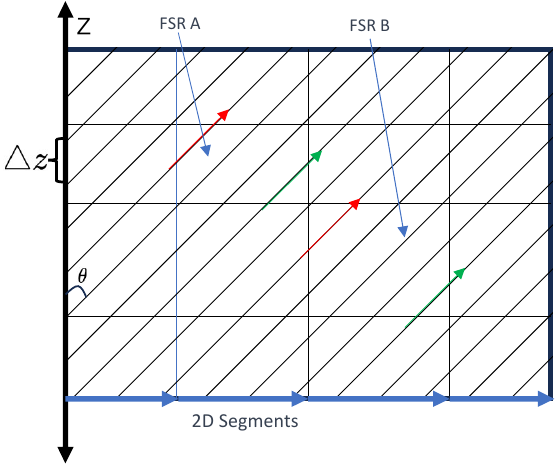}
\caption{Ray Tracing Track Z-Stacks\label{fig1}}
\end{figure}   

It is important to observe that for FSR A as shown in Figure ~\ref{fig1}, there exist several track segments that traverse the full 2D distance of the FSR and, consequently, are identical. The length of their 3D segments ($L_{3D}$) can be straightforwardly calculated as a product of the axial ray spacing ($\bigtriangleup z$) and the number of segments encountered within the axial height ($z_{i}$) of the FSR.

\begin{equation}
L_{3D} =\frac{s_{end}-s_{start}  }{\sin \theta } 
\end{equation}

To leverage this advantage, you can compute the indices of the initial and final tracks that cover the entire length of the 2D segment. To meet this condition, the vertical position of the tracks along the entire 2D segment must be within the range defined by the minimum and maximum axial boundaries of the FSR. Hence, the relevant indices are labeled as $i_{in}$, signifying the track that first crosses the minimum boundary at its lowest point, and $i_{out}$, indicating the track that initially surpasses the maximum boundary at its highest point. These indices can be computed as follows:

\begin{equation}
i_{in}=\left \lceil\frac{z_{min}-min\left ( z_{0}\left ( s_{start}  \right ),z_{0}\left ( s_{end}  \right )     \right )   }{\bigtriangleup z} \right \rceil
\end{equation}

\begin{equation}
i_{out}=\left \lfloor \frac{z_{max}-max\left ( z_{0}\left ( s_{start}  \right ),z_{0}\left ( s_{end}  \right )     \right )   }{\bigtriangleup z}  \right \rfloor
\end{equation}

For each FSR, the track indexes ($i_{in}$ and $i_{out}$) are calculated, in addition to the starting and ending track indexes as given in Equation~\ref{equation6} and Equation~\ref{equation7}. In certain scenarios, especially when the polar angle is sufficiently steep, it is possible for the index $i_{in}$ to be greater than $i_{out}$. This suggests that the FSR can be completely crossed along the axial direction without the need to traverse its entire 2D length. The indices $i_{in}$ and $i_{out}$ can be used to identify these tracks, which share a same 3D segment length $L_{3D}$, as given by:

\begin{equation}
L_{3D} =\frac{z_{max}-z_{min}  }{\cos \theta } 
\end{equation}

The steps outlined above describe the process of generating characteristic rays and segments using the On-The-Fly (OTF) method. Indeed, the On-the-Fly (OTF) method exhibits remarkable performance in reducing memory usage, improving computational efficiency, and handling complex geometries, making it an excellent choice for parallel computing on GPUs. GPUs possess powerful parallel computing capabilities, enabling them to handle massive data and tasks concurrently. The inherent characteristics of the OTF method allow it to fully leverage the parallel computing advantages of GPUs. As a result, the combination of OTF and GPU enables efficient and high-performance computations for a wide range of applications, especially in the field of numerical reactor simulations and other scientific calculations.

\section{Parallel Method based on OTF}
\subsection{Functional Modules}
The computation process can be divided into four main components: input/output, geometric modeling, ray tracing, and iterative solving. The entire program is written in C++ and follows an object-oriented design approach to decompose various functional modules, resulting in a code structure that is easy to extend and maintain. In the porting process, we utilize geometric modeling module and 2D ray tracing module, while migrate the 3D ray tracing and iterative solving modules to the GPU.

\begin{figure}[ht]
\centering
\includegraphics[width=0.8\textwidth]{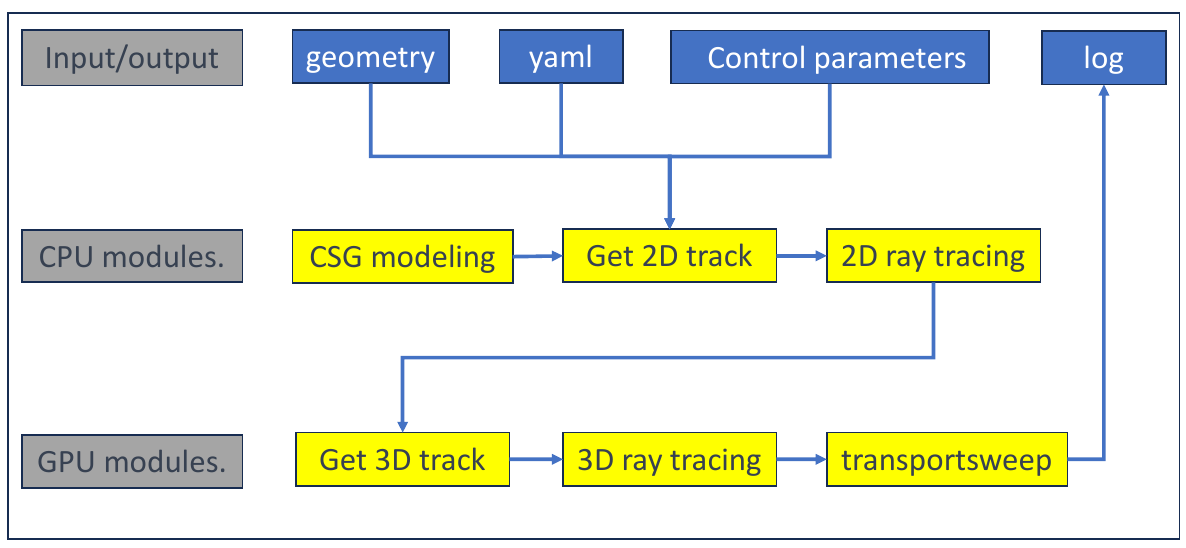}
\caption{Overview of Functional Modules}
\label{fig2}
\end{figure}

Figure~\ref{fig2} illustrates the key functional modules of the ported program and their interrelationships. Initially, the input data of the program is divided into two distinct sections: a geometric model (in YAML format) and runtime parameters (also in YAML format). These runtime parameters encompass the file path for the geometric input and parameters governing the behavior of the control program. Subsequently, the CPU executes 2D ray tracing to generate the requisite data for 3D ray tracing. Ultimately, the necessary data for transport calculations are copied to the GPU, where the GPU invokes the appropriate kernel function to conduct 3D ray tracing and perform transport calculations. The integration of HIPIFY-CLANG during the compilation process enables users to select the desired GPU platform at compile time. Currently available platforms encompass CUDA and HIP, allowing users to tailor their choice according to their computational resources. Employing HIPIFY-CLANG streamlines the compilation process and ensures code optimization for the chosen GPU platform, thereby enhancing performance and efficiency during execution.

\subsection{Code Porting Strategy}

\begin{figure}[ht]
\centering
\includegraphics[width=0.5\textwidth]{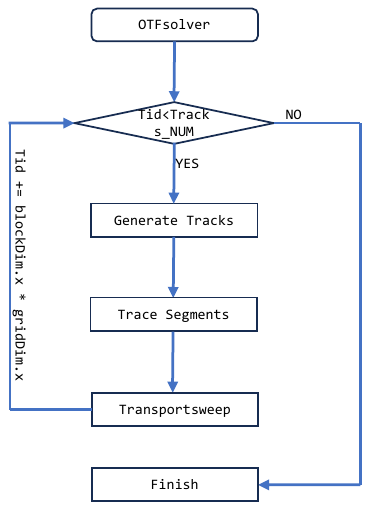}
\caption{Flowchart of Kernel Function}
\label{fig3}
\end{figure}

In the process of porting the code to the GPU, we adopted the OTF method and set its main function as the kernel function, while setting the various functions called by it as device functions. The overall flow of the kernel function is shown in Figure~\ref{fig3}. Firstly, we initialize the number of GPU threads and blocks. This part can be adjusted through an input file, and if not specified, it defaults to 512*512. Next, each thread is assigned a unique tid for indexing the track, where the generation formula for tid is shown in Equation 8. 

\begin{equation}
tid = threadIdx.x + blockIdx.x * blockDim.x
\end{equation}

In the equation, tid represents the unique index of the thread, threadIdx.x represents the index of the thread within its corresponding thread block, blockIdx.x represents the index of the thread block within the grid, and blockDim.x represents the number of threads per thread block. Once a thread completes one iteration, it increases its tid by the step size of the total number of threads and continues to the next iteration until tid exceeds the number of feature lines. Each iteration involves three steps: generating tracks, segmenting the feature lines, and performing calculations on these tracks.

\begin{figure}[ht]
\centering
\includegraphics[width=0.5\textwidth]{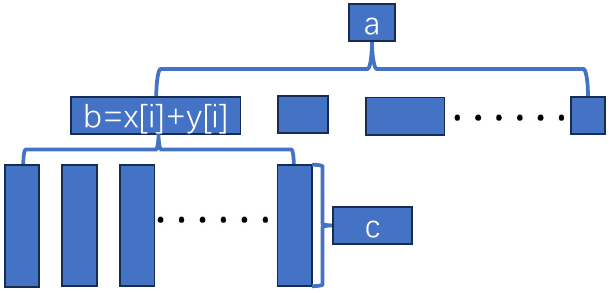}
\caption{struct of Z-STACK}
\label{fig}
\end{figure}

For the OTF method, one of its most pivotal datasets is the Z-STACK. This data structure encapsulates a substantial collection of irregular three-dimensional arrays, as illustrated in Figure 5. To elaborate, a tri-level pointer leads to a secondary pointer of length $a$, and each of these secondary pointers then references a primary pointer of length $b$. The value of $b$ is the result of adding the lengths of two individual one-dimensional arrays, each with a length of $a$. Furthermore, each primary pointer points to a fixed-length array of size $c$. To optimize the efficiency of memory access and enhance the continuity during access operations, we have adopted the strategy of unfolding this three-dimensional array into a one-dimensional array and transferring it to the GPU. During the unfolding process, we meticulously record the positions of each secondary pointer within the unfolded array and store these quantities in an array. Eventually, within the core computational kernel, we employ Equation 8 to perform index computation, thereby enabling effective access to this unfolded one-dimensional array.

\begin{equation}
Z-STACK[i][j][k] = stack[z[i]+j*c+k]
\end{equation}

In the provided equation, $Z-STACK$ represents an irregular three-dimensional array, $stack$ denotes the unfolded one-dimensional array.  $i$, $j$, and $k$ correspond to the indexes within the three-dimensional array. The symbol $z$ represents the marker for each secondary pointer in the two-dimensional array, and $c$ denotes the length of the primary pointer.

To improve program efficiency, we employ several techniques such as storing frequently accessed data in a cache to reduce global memory access, utilizing data prefetching and data alignment to enhance data transfer efficiency. Additionally, we utilize shared memory and registers to store temporary calculation results, minimizing reliance on global memory. Furthermore, we carefully partition data blocks to ensure more contiguous and localized data access, thereby improving memory access efficiency.

\section{Parallel Method based on Tracks}
\subsection{Algorithm Optimization}

\begin{algorithm}
\For{$ext\_id = 0 \to num\_2D\_tracks$}
{
    \For{$p = 0 \to num\_polar$}
    {
           \For{$z = 0 \to tracks\_per\_stack$}
           {
            getTrackOTF\; 
            traceSegmentsOTF\;
            transportsweep\;
           }
    }
}
\caption{OTF Method}
\label{alg1}
\end{algorithm}

The OTF algorithm traversing all tracks in a single pass can be represented using pseudocode as shown in Algorithm~\ref{alg1}. Firstly, for each 2D track, we retrieve the corresponding index of the 3D track. Next, for each polar angle, we perform the next step of the operation. Finally, for each track in the z-stack, we extract the 3D track and obtain its information. Through this pseudocode, it can be observed that by using 2D tracks, polar angles, and z-stacks, we can uniquely locate an independent track.

To achieve efficient computation, we design an indexing coordinate for each 3D characteristic line, which includes its corresponding 2D characteristic line, polar angle, and z-stack information. We pack all the indexing coordinates of the 3D characteristic lines into a one-dimensional array and transfer it to the GPU. On the GPU, each thread looks up the corresponding characteristic line based on its own thread ID (tid) and performs the corresponding calculations. This improvement enables efficient parallel computation on the GPU. The improved GPU parallel pseudocode is illustrated in Algorithm ~\ref{alg2}.

\begin{algorithm}
get thread ID: tid = threadIdx.x + blockIdx.x * blockDim.x\;
\While{tid \textless num\_3D\_tracks}
{
    transportsweeponDevice\;
    tid += blockDim.x * gridDim.x\;
}
\caption{Parallel Method of GPU}
\label{alg2}
\end{algorithm}
\subsection{Memory Optimization}
\begin{figure}
\centering
\includegraphics[width=0.5\textwidth]{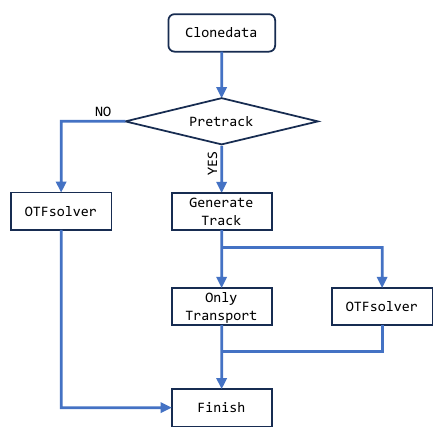}
\caption{Overview of Track Manangment} 
\label{fig4}
\end{figure}

In practical testing, we observed that the OTF method requires very little GPU memory for computing the 3D tracks, resulting in a significant amount of idle GPU memory during computation. Additionally, despite optimizing many conditional statements during program porting, there still remain a substantial number of conditional logic operations during the generation of characteristic lines and segments, impacting computation speed. To address this, we categorized the characteristic lines and stored a portion of them, which have a higher number of segments, in GPU memory continuously, while the remaining lines are generated and computed in real-time. The overall function flow is illustrated in Figure ~\ref{fig4}.

First, preloading of characteristic lines and their corresponding segments is performed, and they are sorted in descending order based on memory usage. Then, the memory usage of characteristic lines is cumulatively added, and when the cumulative value reaches 80\% of the current GPU memory threshold, the process is stopped, and the corresponding indices are recorded. Next, the device function for generating characteristic lines is executed, and these data are stored on the GPU. For this set of characteristic lines, a dedicated kernel function that handles only the transport calculation is directly called. For the remaining characteristic lines, the OTF kernel function is synchronously invoked for computation. Finally, the scalar flux is organized, completing one iteration. The method that employs segment storage is referred to as the EXP method.

This memory management strategy is of great significance in improving efficiency. By preloading and sorting the characteristic lines based on their memory usage, it maximizes the utilization of GPU memory resources, ensuring that the memory usage during computation does not exceed the available memory capacity. Moreover, setting the GPU memory threshold allows for flexible adjustment of the memory usage upper limit to accommodate different-scale computational tasks. This strategy enhances computation efficiency and performance while providing a certain level of user control, enabling customization of the algorithm based on specific requirements.

The aforementioned memory optimization techniques provide an effective solution for characteristic line computation in high-performance computing environments. Such methods have the potential to optimize memory usage and improve computation efficiency, applicable to problem-solving in various scientific and engineering domains. With carefully designed memory management strategies, the computational power of GPUs can be maximized, enhancing computation efficiency and algorithm scalability.

\subsection{Load Balancing}  

The core steps of each computation can be clearly observed from the pseudocode, which are the tracesegment and transportsweep functions. For these two functions, generating each segment requires at least one traversal, and during the transport calculation on each ray, traversing each segment is necessary. Hence, the segment is the smallest computational unit in each iteration.

Meanwhile, our parallel strategy involves mapping the feature lines onto the GPU thread grid. Thus, workload assessment can be based on the number of segments.

To achieve load balancing, we employed an optimization scheme involving preloading based on the number of segments within feature lines. The specific steps are as follows: First, preloading the number of feature lines on the GPU. Then, reorganizing the 3D feature line's coordinate indices. The reorganization entails sorting the feature lines in descending order according to segment count, grouping them in units of 512x512, and alternately reverse sorting every other group. This achieves thread-level load balancing. Lastly, the reorganized feature line indices are employed in the computation.

The effectiveness of these optimization measures will be demonstrated and validated in Chapter 5. Through these steps, we can better harness the parallel computing capabilities of the GPU, achieve load balancing in feature line calculations, enhance computational efficiency, and attain superior performance.

\begin{figure}[!h]
\centering
\includegraphics[width=\textwidth]{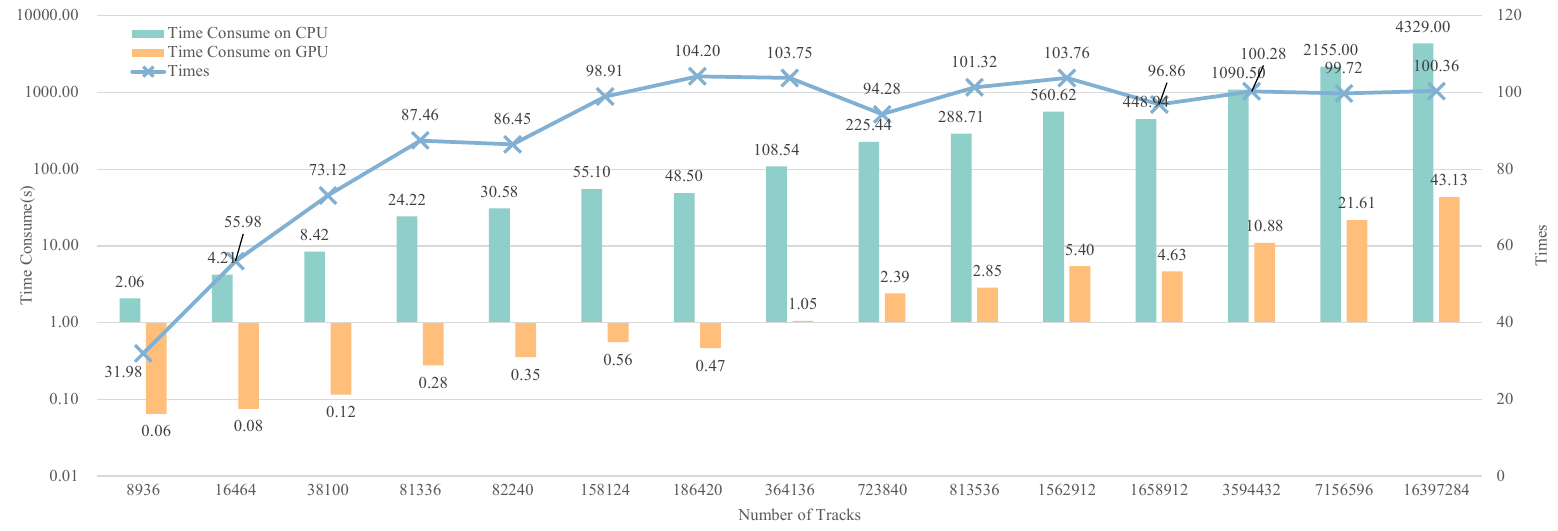}
\caption{Effect of the Accelaration on GPU}
\label{fig5}
\end{figure}
\begin{figure}[!h]
\centering
\includegraphics[width=\textwidth]{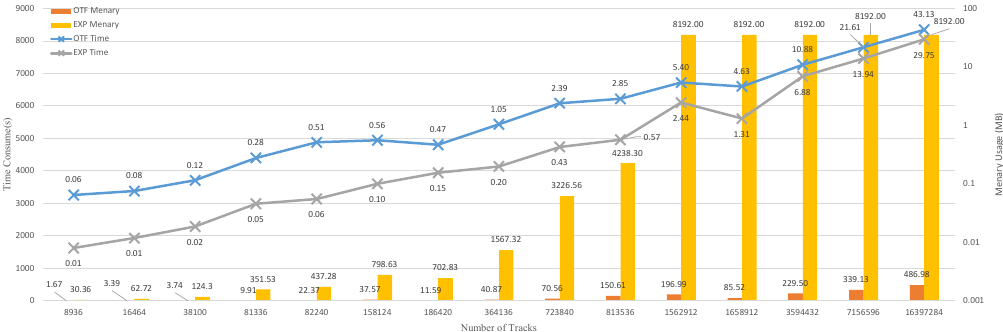}
\caption{Storage and Time Comparison of The Memory Optimization}
\label{fig6}
\end{figure}

\begin{table}
\centering
\caption{Platform}
\label{table1}
\begin{tblr}{
  cells = {c},
  hline{1-2,7} = {-}{},
  hline{6} = {1}{},
}
CPU                   & GPU    & Host Memory \\
                      & 1080Ti &             \\
12700KF               & 2080Ti & 32GB        \\
                      & 3060   &             \\
                      & 4090   &             \\
32-core AMD processor & MI60   &             
\end{tblr}
\end{table}

\section{Results}

In this chapter, we will present the experimental results and data analysis to validate the effectiveness and performance of the GPU parallel method for OTF. We will first verify the correctness of the computation results under different models. Then, we will compare the performance of the ported program with the OTF method in OpenMOC, discussing the results of each experimental setting and evaluation metric in detail. Finally, we will run the program on different GPU platforms to test its compatibility, allowing individual users to choose the GPU platform based on their available computing resources.

\subsection{Correctness Validation}

\begin{table}[!h]
\centering
\caption{Experimental on Correctness Validation}
\label{table2}
\begin{tabular}[width=0.85\linewidth]{@{}ccccc@{}}
\toprule
Tracks & Segments & k-eff    & iterations \\ \midrule
8936            & 1757654            & 1.110546 & 947        \\
16464           & 3642056            & 1.108022 & 1070       \\
38100           & 7382764            & 1.10865  & 1070       \\
81336           & 21025718           & 1.108199 & 736        \\
82240           & 25603240           & 1.108956 & 682        \\
158124          & 46933788           & 1.1099   & 698        \\
186420          & 42456904           & 1.108697 & 736        \\
364136          & 93946642           & 1.111342 & 697        \\
723840          & \num{1.94e8}       & 1.111747 & 689        \\
813536          & \num{2.52e8}       & 1.111872 & 678        \\
1562912         & \num{4.87e8}       & 1.110985 & 685        \\
1658912         & \num{3.91e8}       & 1.112211 & 689        \\
3594432         & \num{9.73e8}       & 1.112428 & 684        \\
7156596         & \num{1.93e9}       & 1.112851 & 684        \\
16397284        & \num{3.88e9}       & 1.113312 & 684        \\ \midrule
\end{tabular}
\end{table}

The C5G7 3D extension benchmark \citep{smith2006benchmark}, released by the Organization for Economic Cooperation and Development Nuclear Energy Agency (OECD/NEA), serves to assess the contemporary deterministic transport methods and codes competency in addressing intricate reactor physics challenges marked by spatial heterogeneities, sans the reliance on spatial homogenization techniques. This benchmark stands as an openly accessible three-dimensional testing ground, offering an intricate portrayal of the multifaceted reactor geometry intricacies within a reasonably expansive computational domain. Multiple whole-core transport codes \citep{gunow2019full,zhang2017calculation,gao2013validation,liu2019material,reda2014calculation,wu2019verification} have been subjected to validation against this benchmark, solidifying its role. In line with this, the tests detailed in this paper are rooted in the C5G7 \citep{humbert2006results,wu2020verification} benchmark case, employing varying numbers of tracks. We conducted tests on multiple platforms with various scenarios involving features of different scales. The configurations for each platform are detailed in Table ~\ref{table1} below. Standard k-eff values and iteration counts \citep{wang2021parallel} were harnessed as yardsticks to gauge the parity of computational outcomes. After conducting extensive cross-platform testing, our findings are summarized in Table ~\ref{table2} below. Importantly, regardless of the platform used to run models of varying scales, our results consistently exhibit uniformity.

These tests were performed to compare the computation results obtained from the ported program with those obtained from the OTF method in OpenMOC. The k-eff values and the number of iterations were used to assess the equality of the computation results.In each parallel test group, the same CPU resources were used, with the only difference being the utilization of GPUs. The test results showed that all fifteen test cases had the same number of iterations during convergence, and the values of k-eff were identical. These test results indicate that the program, when accelerated with GPUs, produces correct computation results.

\subsection{Parallel Testing}

Based on Figure ~\ref{fig5}, we can observe the comparison between CPU and GPU computation results. The GPU computation time is significantly higher than that of the CPU, so we have applied a logarithmic scale on the time axis and indicated the values above the graph. The experiments were conducted on the same computing platform as the correctness verification. Compared to the OpenMOC method\citep{boyd2016parallel}, the GPU-accelerated computation efficiency gradually increased from 30x to 100x, reaching its peak as the computation scale increased. This phenomenon is attributed to the limited utilization of GPU resources due to fewer feature lines at smaller computation scales, which restricts the acceleration effect. With the increase in computation scale, the program can more effectively harness computational resources, resulting in a larger speedup. However, once the speedup reaches 100x, the GPU's performance becomes the bottleneck, and further improvement is not possible.

It is worth noting that the speedup may vary depending on the specific GPU and CPU platforms used. Different hardware architectures and performance characteristics can affect the speedup. Therefore, when using GPU acceleration, it is crucial to consider the specific hardware environment and conduct performance evaluation and optimization accordingly.

In conclusion, GPU acceleration significantly improves the efficiency of numerical reactor calculations, especially for large-scale computations. However, the specific performance may vary depending on the hardware platform, so a comprehensive evaluation and testing are necessary to determine the optimal computational configuration.

Figure ~\ref{fig6} displays the comparison of computational efficiency between the optimized On-the-Fly (OTF) and Explicit (EXP) methods. The results show that the EXP method achieves 1.5 to 10 times higher computational efficiency than the OTF method. This improvement is attributed to the extensive logical inference required by the OTF method during the generation of feature lines and segments, which is not conducive to the GPU architecture. GPUs excel in parallel computations on large-scale datasets. However, the logical inference in the OTF method can introduce computational branches and merges, reducing GPU computational efficiency. As the computation scale increases, the computation time of both methods tends to converge. Therefore, for smaller computation scales or when sufficient GPU resources are available, it is advisable to use the pure EXP method for computation.

\begin{figure}
\centering
\includegraphics[width=0.8\textwidth]{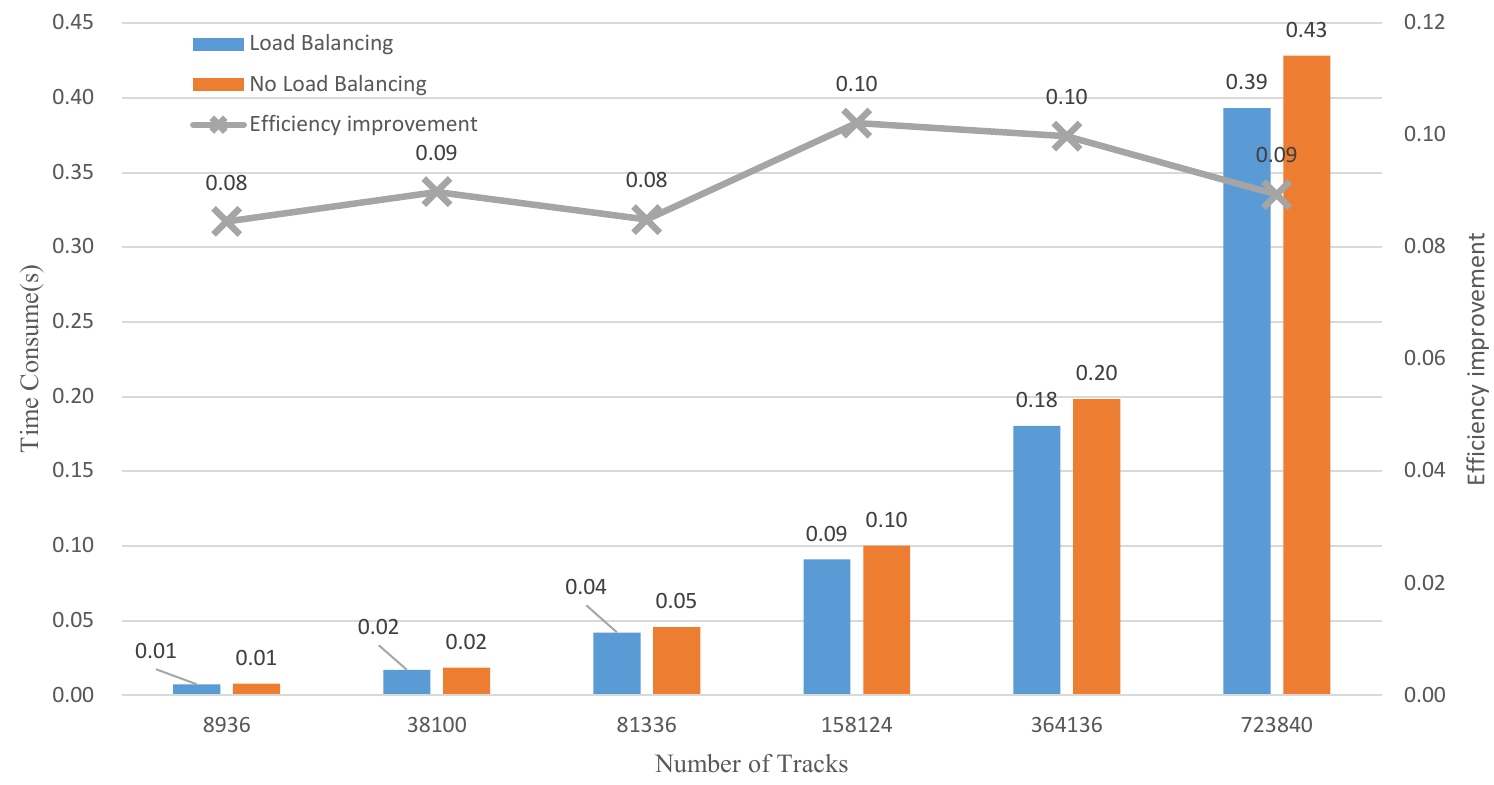}
\caption{load balancing}
\label{fig7}
\end{figure}

Figure~\ref{fig7} provides a visual representation of the comparative analysis between the load-balanced optimization of the On-the-Fly (OTF) method and the original method (without load balancing). The comparison results indicate that the introduction of load-balancing techniques has led to an approximately 10\% improvement in computational efficiency. However, it's important to note that the observed improvements exhibit certain fluctuations, revolving around this central improvement value.These fluctuations arise from the inherent uncertainty in the initial computation order of feature lines, which contributes to the variability in load balancing effectiveness. Despite these fluctuations, it is worth emphasizing that the feature line strategy rearrangement achieved through load balancing consistently brings about noticeable and discernible improvements when contrasted with the original unbalanced approach.By achieving a more uniform distribution of computational tasks among processing resources, the load balancing mechanism optimizes the utilization of available hardware, fostering a higher level of parallelism and concurrency during computation. This consequently leads to the demonstrated improvement in computational efficiency.

\section{Conclusion}
This paper has presented the theoretical foundation, program implementation, and relevant experimental data of a GPU-based parallel three-dimensional MOC. By employing the OTF method for real-time generation and computation, It addresses the deficiency in the field of GPU parallel three-dimensional MOC. Multiple experimental results have demonstrated the accuracy and reliability of the program's computation results, along with a significant improvement in computational efficiency achieved through GPU acceleration. Additionally, the program is adaptable to various GPU platforms, providing users with greater flexibility and choice.

In summary, as a parallel computational program for three-dimensional neutron transport aimed at numerical reactors, this program possesses the capability to handle large-scale parallel computations for various reactor types and high-precision real geometry cases. Future research directions could focus on expanding the program to cover more reactor benchmarks, mathematically modeling the characteristic line generation phase, and representing the computation process using matrices to greatly enhance the program's parallel efficiency on GPUs.

\vspace{6pt}

\bibliographystyle{unsrt}  
\bibliography{references}  






\end{document}